\documentclass[aps,prl,twocolumn,superscriptaddress,showpacs,showkeys]{revtex4}
\usepackage{graphicx}
\usepackage{amsmath,amssymb}

\begin{document}

\title{Evidence for Correlations Between Nuclear Decay Rates and Earth-Sun Distance}


\author{Jere H. Jenkins}
\affiliation{Physics Department, Purdue University, 525 Northwestern Avenue, West Lafayette, Indiana, 47907, USA}

\author{Ephraim Fischbach}
\email[]{ephraim@physics.purdue.edu}
\affiliation{Physics Department, Purdue University, 525 Northwestern Avenue, West Lafayette, Indiana, 47907, USA}

\author{John B. Buncher}
\affiliation{Physics Department, Purdue University, 525 Northwestern Avenue, West Lafayette, Indiana, 47907, USA}

\author{John T. Gruenwald}
\affiliation{Physics Department, Purdue University, 525 Northwestern Avenue, West Lafayette, Indiana, 47907, USA}

\author{Dennis E. Krause}
\affiliation{Physics Department, Purdue University, 525 Northwestern Avenue, West Lafayette, Indiana, 47907, USA}
\affiliation{Physics Department, Wabash College, Crawfordsville, Indiana, 47933, USA}
\author{Joshua J. Mattes}
\affiliation{Physics Department, Purdue University, 525 Northwestern Avenue, West Lafayette, Indiana, 47907, USA}


\date{\today}

\begin{abstract}
Unexplained periodic fluctuations in the decay rates of $^{32}$Si and $^{226}$Ra have been reported by groups at Brookhaven National Laboratory ($^{32}$Si), and at the Physikalisch-Technische-Bundesandstalt in Germany ($^{226}$Ra).  We show from an analysis of the raw data in these experiments that the observed fluctuations are strongly correlated in time, not only with each other, but also with the distance between the Earth and the Sun.  Some implications of these results are also discussed, including the suggestion that discrepancies in published half-life determinations for these and other nuclides may be attributable in part to differences in solar activity during the course of the various experiments, or to seasonal variations in fundamental constants.
\end{abstract}

%

\maketitle


Following the discovery of radioactivity by Becquerel in 1896 \cite{Bequerel:1896} an intense effort 
was mounted to ascertain whether the decay rates of nuclides could be affected by 
external influences including temperature, pressure, chemical composition, 
concentration, and magnetic fields.  By 1930, Rutherford, Chadwick, and 
Ellis \cite[p.~167]{Rutherford:1930} concluded that \textquotedblleft{}The rate of transformation of an element has been 
found to be a constant under all conditions.\textquotedblright{}  (For decays resulting from 
K-capture, or for beta-decays in strong ambient electromagnetic fields, 
the situation is slightly more complicated, since these decays are influenced 
by the electron wave functions which can be affected by external pressure or 
fields \cite{Fassio:1969,Hahn:1976,Ohtsuki:2004}.)   For $^{32}$Si and $^{226}$Ra, 
which decay by beta- and alpha-emission, 
respectively, fluctuations in the counting rates (in the absence of strong 
external electromagnetic fields) should thus be uncorrelated with any external time-dependent 
signal, as well as with each other.  In what follows we show that neither of 
these expectations is realized in data we have analyzed for $^{32}$Si and $^{226}$Ra, 
thus suggesting that these decays are in fact being modulated by an external influence.

Between 1982 and 1986, Alburger, et al.~\cite{Alburger:1986} measured the half-life of $^{32}$Si at 
Brookhaven National Laboratory (BNL) via a direct measurement of the counting 
rate as a function of time.  If \begin{math}N(t){}\end{math} denotes the number of surviving atoms 
starting from an initial population \begin{math}N_0{}\end{math} at \begin{math}t=0{}\end{math}, then the familiar exponential 
decay law, \begin{math}N(t)=N_0e^{-\lambda t}\end{math}, 
leads to \begin{math}\dot{N}\equiv{}dN/dt=- \lambda N_0 e^{- \lambda t}\end{math}
where \begin{math}\lambda =\ln(2)/T_{1/2}\end{math}.  A plot of \begin{math}\ln\left[\dot{N}(t)\right]\end{math}  as a function of 
time is then a straight line whose slope is \begin{math}\lambda{}\end{math}, which then gives the half-life \begin{math}T_{1/2}{}\end{math}.  At the time this experiment was initiated, the $^{32}$Si half-life was believed 
to be in the range of \begin{math} 60 \lesssim T_{1/2} \lesssim 700\end{math} yr, and hence a multi-year counting 
experiment was needed to obtain a measureable slope.  As in other counting 
experiments, the counting rate for $^{32}$Si was continually monitored in the same 
detector against a long-lived comparison standard, which in the BNL experiment 
was $^{36}$Cl (\begin{math}T_{1/2}{}\end{math}=301,000 yr).  Since the fractional change in the $^{36}$Cl counting 
rate over the four year duration of the experiment was only \begin{math}O{}(10^{-5})\end{math}, which was 
considerably smaller than the overall uncertainty of the final result, 
$T_{1/2} (^{32}$Si)=172(4) yr, the $^{36}$Cl decay rate was assumed to be constant.  Any time 
dependence for $^{36}$Cl beyond the expected  statistical fluctuations was 
then presumed to arise from various systematic effects, such as drift in the 
electronics.  By computing the ratio \begin{math}^{32}\mbox{Si}/^{36}\mbox{Cl}\equiv\dot{N}(^{32}\mbox{Si})/\dot{N}(^{36}\mbox{Cl})\end{math}, these apparatus-dependent systematic 
effects should have largely cancelled, and hence this ratio was used to obtain 
the half-life of $^{32}$Si.  On the other hand, barring an accidental cancellation, 
time-dependent contributions to the $^{32}$Si and $^{36}$Cl decay rates themselves would 
not cancel in the ratio $^{32}$Si/$^{36}$Cl.

\begin{figure}[h]
 \includegraphics[height=60mm] {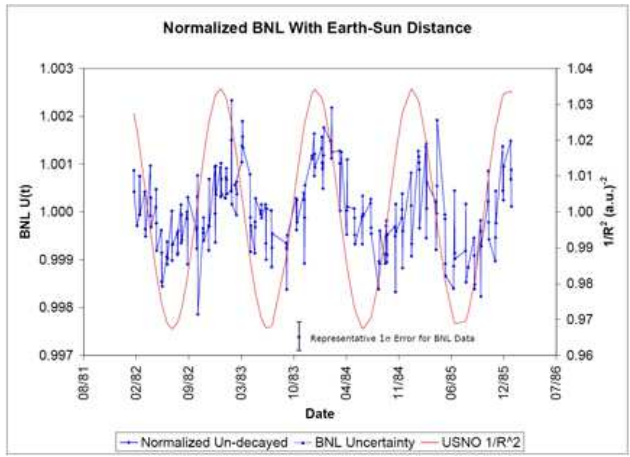}
 \caption{\label{fig:Fig1_BNL-Raw} Plot of $U(t)$ for the raw BNL $^{32}$Si/$^36$Cl ratio along with $1/R^2$ where $R$ is the Earth-Sun distance in units of $1/(\mbox{a.u.})^2$.  $U(t)$ is obtained by multiplying each data point by $\exp(+ \lambda t)$ where $\lambda =\ln(2)/T_{1/2}$ and $T_{1/2}$=172 yr for $^{32}$Si.  The left axis gives the scale for the normalized $U(t)$, and the right axis denotes the values of $1/R^2$ in units of $1/(\mbox{a.u.})^2$ obtained from the U.S. Naval Observatory (USNO).  The fractional change in $^{32}$Si counting rates between perihelion and aphelion is approximately $3\times 10^{-3}$.  As noted in the text, the correlation coefficient between the BNL data and $1/R^2$ is $r$=0.52 for $N$=239 points.  The formal probability that the indicated correlation could have arisen from uncorrelated data sets is $6\times 10^{-18}$.}
 \end{figure}

\begin{figure}[h]
 \includegraphics[height=60mm]{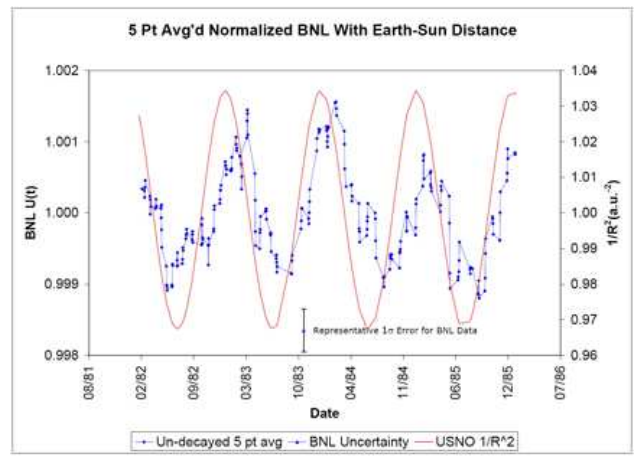}
 \caption{\label{fig:Fig2_BNL5Pt} Plot of the 5 point rolling average of $U(t)$ for the BNL $^{32}$Si data shown in Figure 1.  
 Each data point represents the average of 5-points centered on the original datum, which serves to smooth short term fluctuations in the 
 $^{32}$Si/$^{36}$Cl ratio arising from influences other than a possible annual $1/R^2$ variation.  
 As noted in the text, the correlation coefficient between the BNL data and $1/R^2$ is $r$=0.65 for $N$=235 points.  The formal probability that the indicated correlation could have arisen from uncorrelated data sets is $1\times 10^{-29}$. }
 \end{figure}

The BNL data for the ratio $^{32}$Si/$^{36}$Cl revealed an unexpected annual variation 
of $^{32}$Si/$^{36}$Cl which could not be accounted for by the known effects of temperature, 
humidity, or pressure variations in their detector\cite{Alburger:1986}.  We obtained the raw data from 
the BNL experiment in conjunction with an independent effort to apply a new randomness 
test \cite{Tu:2003,Tu:2005} to nuclear decays, and the BNL data are shown in Fig.~\ref{fig:Fig1_BNL-Raw}.  When comparing the results from experiments on different nuclides, it is convenient to study the function  
\begin{math}U(t)\equiv \left[\dot{N}(t)/\dot{N}(0)\right]\exp(+\lambda t)\end{math}
rather than $\dot{N}(t)$  itself, since $U(t)$ should be time-independent for all nuclides.  For $^{32}$Si, 
we used $\lambda =4.0299 \times 10^{-3} \mbox{yr} ^{-1}$  from Ref.~\cite{Alburger:1986}.  Figure \ref{fig:Fig1_BNL-Raw} exhibits $U(t)$ for 
the $^{32}$Si/$^{36}$Cl BNL data, along with a 
plot of $1/R^2$, where $R$ is the distance between the Earth and the Sun.  An annual 
modulation of the $^{32}$Si/$^{36}$Cl ratio is clearly evident, as was first reported in Ref.~\cite{Alburger:1986}.  
The Pearson correlation coefficient, $r$, between the raw BNL data and $1/R^2$ is $r$=0.52 
for $N$=239 data points, which translates to a formal probability of $6 \times 10^{-18}$ that this 
correlation would arise from two data sets which were uncorrelated.  As shown in 
Figure \ref{fig:Fig2_BNL5Pt}, the correlation coefficient increases to $r$=0.65 for $N$=235 data points when 
a 5 point rolling average is applied.  There is also a suggestion in 
Figs.~\ref{fig:Fig1_BNL-Raw} and \ref{fig:Fig2_BNL5Pt} of a phase 
shift between $1/R^2$ and the BNL data, which we discuss in greater detail below.

\begin{figure}[h]
 \includegraphics[height=60mm]{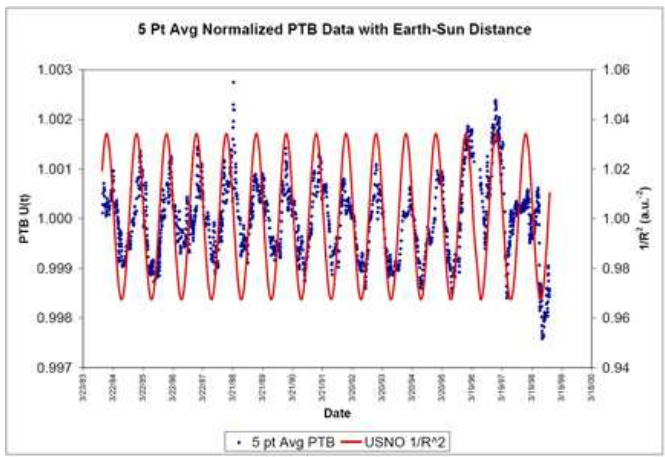}
 \caption{\label{fig:Fig3_PTB5Pt} Plot of $U(t)$ for the PTB $^{226}$Ra data along with $1/R^2$, where $R$ is the Earth-Sun distance.  See caption to Fig. 1 for further details.  The fractional change in the $^{226}$Ra counting rates between perihelion and aphelion is approximately $3\times10^{-3}$.  As noted in the text, the correlation coefficient between the PTB data and $1/R^2$ is $r$=0.66 for $N$=1968 points.  The formal probability that the indicated correlation could have arisen from uncorrelated data sets is $2\times10^{-246}$.  Note that the $1\sigma$ error bars for the PTB data lie within the data points themselves. }
 \end{figure}

The strong correlation between the BNL decay data and the annual modulation of the 
Earth-Sun distance suggests that the $^{32}$Si/$^{36}$Cl ratio may be responding to some 
influence originating from the Sun.  If this is indeed the case, then the effects of 
this influence would be expected to be present in other decays as well.  Although 
there are hundreds of potentially useful nuclides whose half-lives have been measured, 
the data from many of the experiments we examined were generally not useful, most 
often because data were not acquired continuously over sufficiently long time periods.  
However, we were able to obtain the raw data from an experiment carried out at the 
Physikalisch-Technische Bundesandstalt (PTB) in Germany \cite{Schrader:2008,Siegert:1998} measuring the half-life 
for $^{152}$Eu, in which $^{226}$Ra was the long-lived comparison standard.  This experiment, 
which extended over 15 years, overlapped in time with the BNL experiment for 
approximately 2 years, and exhibited annual fluctuations in the $^{226}$Ra data similar 
to those seen at BNL.  Figure \ref{fig:Fig3_PTB5Pt} exhibits the PTB data as a 5 point rolling average, 
and it is evident from the figure that the PTB data closely track the annual 
variation of $1/R^2$.  The Pearson correlation coefficient $r$ for the data in Fig.~\ref{fig:Fig3_PTB5Pt} is $r$=0.66 
for $N$=1968 data points, corresponding to a formal probability of $2 \times 10^{-246}$ that this 
correlation could arise from two data sets which were uncorrelated.  As in the case 
of the BNL data, there is also a suggestion of a phase shift between $1/R^2$ and the 
PTB data (see below), although this phase shift appears to be smaller than for the BNL data.

\begin{figure}[h]
 \includegraphics[height=55mm]{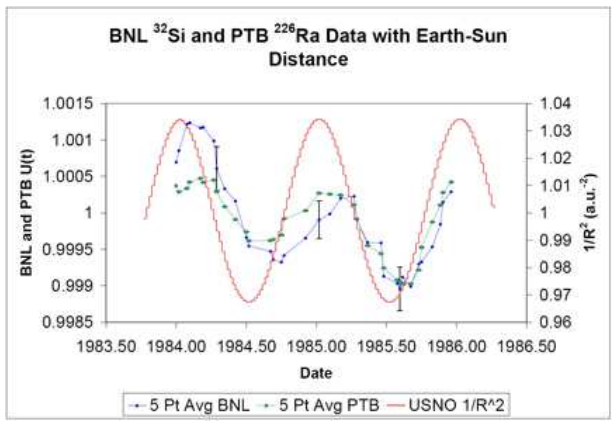}
 \caption{\label{fig:Fig4_BNL-PTB} Correlation between the decay rates of $^{32}$Si at BNL and $^{226}$Ra at PTB.  The BNL and PTB data for $U(t$) have been averaged in common weekly bins for purposes of comparison.  The correlation coefficient between the BNL and PTB data is $r$=0.88, which corresponds to a probability of $4\times10^{-12}$ that the BNL/PTB correlation could have arisen from uncorrelated data sets as a result of statistical fluctuation.  Error bars are shown for representative BNL data points, and the error bars for the PTB data lie within the points themselves. }
\end{figure}

Since the BNL and PTB data each exhibit strong correlations with the annual variation 
of $1/R^2$, it is not surprising that these data correlate with each other.  Figure \ref{fig:Fig4_BNL-PTB} 
exhibits this correlation along with the annual variation of $1/R^2$.  The Pearson 
correlation coefficient for the BNL and PTB data is $r$=0.88 for $N$=35 points, which 
corresponds to a formal probability of $4 \times 10^{-12}$ that this correlation could have 
arisen from two uncorrelated data sets.  Moreover, the difference in latitude 
between BNL and PTB, as well as the difference in their climates, argues against 
an explanation of this correlation in terms of seasonal variations of climatic 
conditions such as temperature, pressure, and humidity etc., which could have 
influenced the respective detection systems.  As an example, radon concentrations 
are known to fluctuate seasonally, as has been noted in Ref.~\cite{Siegert:1998}, and it was suggested 
that the decay of $^{222}$Rn could lead to a seasonally dependent charge distribution on the 
experimental apparatus.  However, this effect is extremely small given the low 
counting rates that typically arise from radon background \cite{Wissman:2006}, and in any case, 
the PTB data shown in Fig.~\ref{fig:Fig3_PTB5Pt} were corrected for background.

The preceding considerations, along with the correlations evident in Fig.~\ref{fig:Fig4_BNL-PTB}, suggest 
that the time-dependence of the $^{32}$Si/$^{36}$Cl ratio and the $^{226}$Ra decay rate are being 
modulated by an annually varying flux or field originating from the Sun, although 
they do not specify what this flux or field might be.  The fact that the two decay 
processes are very different (alpha decay for $^{226}$Ra and beta decay for $^{32}$Si) would 
seem to preclude a common mechanism for both.  However, recent work by Barrow and 
Shaw \cite{Barrow:2008,Shaw:2007} provides an example of a type of theory in which the Sun could affect 
both the alpha- and beta-decay rates of terrestrial nuclei.  In their theory, the 
Sun produces a scalar field $\phi$ which would modulate the terrestrial value of the 
electromagnetic fine structure constant $\alpha_{EM}$.  This could, among other effects, 
lead to a seasonal variation in alpha and beta decay rates, both of which are 
sensitive to $\alpha_{EM}$ \cite{Uzan:2003}.  We note from Fig. \ref{fig:Fig3_PTB5Pt} that the fractional difference 
between the $^{226}$Ra counting rates at perihelion and aphelion is $\approx3\times 10^{-3}$, and
this would require that the coupling constant $k_{\alpha}$ of $\phi$ to $\alpha_{EM}$ should be
$k_{\alpha} \approx 3 \times10^{6}$.  However, this is substantially larger than the value
$k_{\alpha}=(-5.4\pm5.1)\times10^{-8}$ inferred from a recent trapped ion experiment \cite{Barrow:2008,Rosenband:2008}
Although the specific model of Refs.~\cite{Barrow:2008,Shaw:2007,Uzan:2003} 
would not account 
for the $^{32}$Si and $^{226}$Ra data quantitatively, variants of this model might work. This includes
models in which separate scalar fields $\phi_1$ and $\phi_2$ couple, respectively, to $\alpha_{EM}$
and to the electron-proton mass ratio $m_e/m_p$.

Another interesting possibility is that terrestrial radioactive nuclei are 
interacting in a novel way with the neutrino flux $\Phi_{\nu}$ emitted from the interior 
of the Sun.  This flux also varies with $1/R^2$, and the resulting seasonal modulation 
of $\Phi_{\nu}$ has been observed by Super-Kamiokande \cite{Hosaka:2006,Yoo:2003}.  This possibility is supported by 
the data we report in Ref.~\cite{Jenkins:2008} in which we present evidence for the 
possible detection of a change in the decay rate of $^{54}$Mn during the solar flare 
of 13 December 2006.  As noted in Ref. \cite{Jenkins:2008}, the coincidence in time between the 
change in the $^{54}$Mn counting rate and the solar flare, along with other observations, 
is consistent with a mechanism based on a change in $\Phi_{\nu}$ during the solar flare.

We note that irrespective of the origin of the solar flare data, or of the 
correlations evident in Figs.~\ref{fig:Fig1_BNL-Raw}-\ref{fig:Fig4_BNL-PTB}, 
the existence of these effects may explain 
discrepancies in various half-life determinations reported in the literature.  
Examples are $^{32}$Si, $^{44}$Ti and $^{137}$Cs, among many others \cite{Alburger:1986,Alburger:1990,Ahmad:1998,Woods:1990}.  If nuclides 
such as $^{32}$Si, $^{36}$Cl, and $^{226}$Ra respond to changes in the solar neutrino flux 
due to the time-dependence of $1/R^2$, then they can also respond to changes in 
intrinsic solar activity which are known to occur over time scales both longer 
and shorter than one year.  Thus, depending on when half-life measurements 
were made, and on the specific techniques employed, it is possible that some 
of the half-life discrepancies reported in the literature could be reconciled 
if appropriate data on solar activity become available.

Returning to Fig.~\ref{fig:Fig4_BNL-PTB}, we briefly explore the suggestion noted above of a 
possible phase shift of $1/R^2$ relative to both the BNL and PTB data.  Although 
this may be an experimental artifact arising from binning effects, etc., such 
a phase shift could also arise from other smaller contributions to periodic 
variations in neutrino flux.  Possibilities for such contributions were 
explored in Ref.~\cite{Yoo:2003}, where a search was made for short time variations in the 
observed flux at Super-Kamiokande arising from either the 7.25$^\circ$ inclination 
of the solar axis relative to the ecliptic, or from fluctuations in the 
temperature of the solar core.  A modulation of the neutrino flux arising from 
a coupling between a neutrino magnetic moment and a latitudinally inhomogeneous 
solar magnetic field \cite{Sturrock:1998} could also account for a possible phase shift.  Although 
there is no compelling evidence at present for such short time variations at 
Super-Kamiokande\cite{Hosaka:2006,Yoo:2003}, the statistical power of the BNL, PTB, and similar data 
sets may prove to be a useful tool in the search for such effects.  Yet another 
possible explanation for the apparent phase shift could be a seasonally-varying 
velocity-dependent effect similar to that observed by the DAMA/LIBRA collaboration \cite{Bernabei:2008}.

In summary, we have presented evidence for a correlation between changes in 
nuclear decay rates and the Earth-Sun distance.  While the mechanism responsible
for this phenomenon is unknown, theories involving variations in fundamental constants
could give rise to such effects.  These results are also consistent with the 
correlation between nuclear decay rates and solar activity suggested by 
Jenkins and Fischbach \cite{Jenkins:2008} if the latter effect is interpreted as possibly arising 
from a change in the solar neutrino flux. These conclusions can be tested in a 
number of ways.  In addition to repeating long-term decay measurements on Earth, 
measurements on radioactive samples carried aboard spacecraft to other planets 
would be very useful since the sample-Sun distance would then vary over a much 
wider range.  The neutrino flux hypothesis might also be tested using samples 
placed in the neutrino flux produced by nuclear reactors.

\begin{acknowledgments}
The authors are deeply indebted to D. Alburger and G. Harbottle for supplying us with the raw data from the BNL experiment, and to H. Schrader for providing to us the raw PTB data.  We also with to thank B. Budick, B. Craig, M. Fischbach, V. Flambaum, A.M. Hall, A. Longman, E. Merritt, T. Mohsinally, D. Mundy, J. Newport, B. Revis, and J. Schweitzer for their many contributions to this effort.  The work of E.F. was supported in part by the U.S. Department of Energy under Contract No. DE-AC02-76ER071428.
\end{acknowledgments}
\bibliography{ptb-bnl}

\end{document}